\documentstyle [12pt]{article}
\def\1{\mbox{I\hspace{-.15em}1}}

\def\R{{\rm I\hspace{-.15em}R}}

\def\C{\hspace{3pt}{\rm l\hspace{-.47em}C}}

\def\b{\begin{equation}}
\def\e{\end{equation}}
\def\bee{\begin{enumerate}}
\def\eee{\end{enumerate}}
\title{ ``Massive'' vector field in de Sitter space}
\author{J-P. Gazeau$^1$\thanks{e-mail: gazeau@ccr.jussieu.fr}, M.V.
Takook$^{1,2}$\thanks{e-mail: takook@ccr.jussieu.fr}}
\date{\today}

\begin{document}

\maketitle {\it \centerline{$^1$ Laboratoire de Physique
Th\'eorique de la Mati\`ere Condens\'ee} \centerline{ Universit\'e
Paris $7$ Denis-Diderot,75251 Paris Cedex 05, FRANCE}
\centerline{$^2$  Department of Physics, Razi University,
Kermanshah, IRAN} }
\begin{abstract}

We present in this paper a covariant
quantization of the ``massive''  vector field on de Sitter (dS)
space based on analyticity in the complexified pseudo-Riemanian
manifold. The correspondence between unitary
irreducible representations of the de Sitter group and the field
theory on de Sitter space-time is essential in our approach. We
introduce the
Wightman-G\"arding axiomatic for vector field on dS space. The Hilbert
space structure and the unsmeared  field operators
$K_\alpha(x)$ are also defined. This work is in the direct continuation
of  
previous one concerning the scalar and the spinor cases.
\end{abstract}

\vspace{0.5cm}
{\it Proposed PACS numbers}: 04.62.+v, 03.70+k, 11.10.Cd, 98.80.H
\vspace{0.5cm} 

\section{Introduction}

In two previous papers, the various physical
motivations for studying the quantum field in de Sitter space were
explained [Gazeau and al,
$1999$-a,b]. In dS space-time, familiar Minkowskian or galilean physical
quantities 
like mass
or energy cannot be envisaged in a clear operational way. A time-like
Killing vector field
cannot be globally defined, and we cannot deal with
a four-momentum
$p^\mu$ that satisfies $p^2=$constant. On the other hand, there is a
five-component vector
$(\xi^\alpha)= (\xi^0,\vec \xi, \xi^4)$ with $\xi^2=0$, which is similar
to the 
$p^\mu$ in the null curvature limit. Again, a precise protocole of
measurement of such purely
de-Sitterian quantities is still lacking.

Yet,  the principle of causality is well
defined [B\"orner, D\"urr, $1969$; Bros, Moschella, $1996$] in de Sitter
space.
A field called ``massive'' propagates inside
the light-cone and corresponds to a
massive Poincar\'e field in the null curvature limit. We call a field
``massless'' if it
propagates on the dS light-cone and if it
corresponds to a massless Poincar\'e field at $H=0$. Only de Sitter 
vector fields of the
``massive'' type will be considered in this paper. In the case of a
``massless'' dS vector field ( dS QED for instance), we have to resort
to 
 quantization {\it \`a la } Gupta-Bleuler in order to obtain a covariant
construction. This
question  will be addressed in a forthcoming paper.

The field equations for the scalar, spinor, and vector fields in dS
space were established by
Dirac [Dirac,
$1935$]. The solution to the latter case was presented by B\"orner and
D\"urr [B\"orner, D\"urr,
$1969$; Schomblond, Spindel, $1976$] in terms of flat coordinates,
which covers only one half of the dS hyperboloid. In $1986$,
Allen calculated the vector two-point functions in terms of the
geodesic distance. The latter is independent of the choice of coordinate
system [Allen, $1986$]. We present in this paper the Hilbert space
structure and the vector field operator in terms of
coordinate-independent 
dS plane waves. The construction is
based on analyticity properties offered by the complexified
pseudo-Riemanian manifold in which dS 
manifold is embedded, and we refer to [Bros, Moschella, $1996$] 
for a comprehensive review of these rigourous  
mathematical results concerning the functional analysis side of QFT. Our
present aim is rather
to make explicit the extra algebraic structure 
inherent to the vector case.

In Section 2 we describe the dS-vector field equation as an eigenvalue
equation of the Casimir
operator. We define the notations and introduce the two independent
Casimir
operators. In his Thesis dissertation, Takahashi [Takahashi, $1963$]
employed 
two parameters $p$ and
$q$ for characterizing the representations of the dS group. These
parameters behave like a spin ($s$) and a mass ($m$) in the Minkowskian
limit. 
In Section 3 we solve the field equation. The solution is written in
terms of a scalar field
$\phi
$ and a five-component generalized ``polarization" vector ${\cal E}$
 $$ K (x)= {\cal E}(x,\xi)\phi (x). $$
In contrast to the Minkowskian situation, the vector ${\cal E}(x,\xi)$
is a
function of the space-time point  $x^\alpha$. This is due to the fact
that the momentum
operators acquire a spin part [B\"orner, D\"urr, $1969$]. This
five-component vector ${\cal
E}(x,\xi)$ is precisely defined in order to obtain 
the usual polarization vector
at the Minkowskian limit $H=0$. These solutions are not globally defined
due to the
presence of a multiform phase factor. The solution extended to the
complex dS space 
can actually be
considered for solving this problem [Bros, Moschella,
$1996$].

In Section 4, we define the Wigthman two-point functions (${\cal
W}_{\alpha \alpha'}(x,x')$), which satisfies the conditions of: a)
positiveness, b) locality, c) covariance, d) normal analyticity, e)
transversality and f) divergencelessness. The normal analyticity allows
one to define this Wigthman two-point function ${\cal W}_{\alpha
\alpha'}(x,x')$ as the boundary value of an analytic two-point function
$W_{\alpha \alpha'}(z,z')$ from the tube domains. The normal
analyticity is related to the Hadamard condition which selects an
unique vacuum state in dS space [Allen, $1985$]. $W_{\alpha
\alpha'}(z,z')$ is defined in terms of dS plane-waves in their
tube domains. Then, the Hilbert space structure is introduced and
the field operator $K (f)$ is defined. We also give a
coordinate-independent formula for the unsmeared field operator
$K(x)$.

A brief conclusion and outlook are given in Section 5. In that
part, it is concluded that the ``massless'' vector field treatment
requires
an indecomposable
representation of dS group and the construction of the corresponding
covariant quantum field.

\section{dS-vector field equation}

The de Sitter space is an elementary solution of the cosmological
Einstein equation.
It is conveniently seen as a hyperboloid embedded in a five-dimensional
Minkowski space
     \b X_H=\{x \in \R^5 ;x^2=\eta_{\alpha\beta} x^\alpha x^\beta
=-H^{-2}\},\;\;
      \alpha,\beta=0,1,2,3,4, \e
where $\eta_{\alpha\beta}=$diag$(1,-1,-1,-1,-1)$. The kinematical
group of the de Sitter space is the $10$-parameter group
$SO_0(1,4)$ and its contraction limit $H=0$ is the
Poincar\'e group. There are two Casimir operators
    $$ Q^{(1)}=-\frac{1}{2}L_{\alpha\beta} L^{\alpha\beta} $$
     \b
Q^{(2)}=-W_{\alpha}W^{\alpha},\;\;\;W_{\alpha}=-\frac{1}{8}\epsilon_
  {\alpha\beta\gamma\delta\eta}L^{\beta\gamma} L^{\delta\eta}, \e
where the $L_{\alpha\beta}$'s are the infinitesimal generators and
$\epsilon_ {\alpha\beta\gamma\delta\eta}$ is the usual
antisymmetrical tensor. The de Sitter metrics reads
$$ds^2=\eta_{\alpha\beta}dx^{\alpha}dx^{\beta}=g_{\mu\nu}^{dS}dX^{\mu}dX^{\nu},\;\;
\mu=0,1,2,3,$$ 
where the $X^\mu$'s are the $4$ space-time coordinates in
dS hyperboloid. Different coordinate systems can be chosen [Mottola,
$1985$]. 
The wave equation for the vector field
$A_\mu(X)$ propagating on de Sitter space can be derived from
a variational principle using the action integral $(\hbar=1)$ [Allen,
$1986$]
     \b S(A)=\int_{M_H}
(\frac{1}{4}F^{\mu\nu}F_{\mu\nu}+\frac{1}{2}m_H^2A^{\mu}A_{\mu})d\sigma,\e
where $F^{\mu\nu}=\nabla_\mu A_\nu-\nabla_\nu A_\mu,\;\;m_H$ is a
``mass",  and $d\sigma$ is  the $O(1,4)$-invariant measure on $M_H$.
The variational principle applied to $(2.3) $ gives the field
equation
  \b \nabla_\mu F^{\mu\nu}+m_H^2A^\nu=\nabla_\mu (\nabla^\mu
A^\nu-\nabla^\nu A^\mu)+m_H^2A^\nu=0.\e
The antisymmetry of $F^{\mu\nu}$ implies $m_H^2\nabla\cdot A=0$ [Allen,
$1986$]. In the case of the ``massive'' vector field, $m_H\neq 0$ and
we have \b \nabla\cdot A=0\e.
Therefore the wave equation is
    \b (\Box_H +3H^2+m_H^2) A_\mu(X)=0.\e
The five-component vector field notation
$K_{\alpha}(x)$ is used in the following discussion. With this  notation
we can clarify the
relation between the field and the unitary irreducible representations
(UIR)
of the dS group. It is also simpler to express the solution in
terms of the scalar field. The four-component vector field
$A_{\mu}(X)$ is locally determined by a five-component vector
field $K_{\alpha}(x)$ through the relation
     \b A_{\mu}(X)=\frac{\partial x^{\alpha}}{\partial
            X^{\mu}}K_{\alpha}(x(X)),\;\; K_{\alpha}(x)=
            \frac{\partial X^{\mu}}{\partial x^{\alpha}} A_{\mu}(X(x)).
\e
This five-component vector field quantity has to be viewed as an
homogeneous function of the
$\R^5$-variables $x^{\alpha}$ with some arbitrarily chosen degree
$\sigma$
        \b x^{\alpha}\frac{\partial }{\partial
x^{\alpha}}K_{\beta}(x)=x\cdot \partial K_\beta
(x)=\sigma
        K_{\beta}(x). \e
It also satisfies the condition of transversality [Dirac, $1935$]
        \b x\cdot K(x)=0. \e
The wave equation satisfied by $K$ can be established in terms of the
tangential (or transverse)
derivative
$\bar \partial$ on de Sitter space
        \b \bar \partial_\alpha=\theta_{\alpha
\beta}\partial^\beta=\partial_\alpha
         +H^2x_\alpha x\cdot \partial,\;\;\;x\cdot \bar \partial=0,\e
where $\theta_{\alpha \beta}=\eta_{\alpha \beta}+H^2x_{\alpha}x_{
\beta}$ is  the transverse projector.  $K$ corresponds to $A$
through $(2.7)$, so we have $$\nabla_\mu A_\nu \longrightarrow
\theta _\alpha^{\alpha '}\theta _\beta^{\beta '}\partial_{\alpha
'}K_{\beta '}.$$
 Hence the field equation reads:
  \b (H^{-2}(\bar \partial)^2+2)K(x)-2x\bar
\partial\cdot K(x)+ H^{-2}\bar \partial
\partial\cdot K+H^{-2}m_H^2K=0,  \e
which, thanks to $(2.9)$ and divergenceless condition $\partial\cdot
K=0$, simplifies to
 \b (H^{-2}(\bar
\partial)^2+2+H^{-2}m_H^2)K(x)=0.  \e In terms of Laplace-Beltrami
operator on de Sitter space, $ -H^2\Box_H=Q_0=-H^2(\bar
\partial)^2$, we obtain \b
(Q_0-2-H^{-2}m_H^2)K(x)=0=(\Box_H+2H^2+m_H^2)K(x).  \e 
Let us now make the things more precise in the context of representation
theory. The equation
$(2.13)$ has indeed a clear group-theoretical content. The Casimir
operator $Q_1^{(1)}$ is
defined by \b    Q_1^{(1)}=-\frac{1}{2}
L^{\alpha\beta}L_{\alpha\beta}=-\frac{1}{2}
(M^{\alpha\beta}+S^{\alpha\beta})(M_{\alpha\beta}+S_{\alpha\beta}),\e
where $M_{\alpha\beta}=-i (x_\alpha \partial_\beta-x_\beta
\partial_\alpha)=-i (x_\alpha \bar \partial_\beta-x_\beta \bar
\partial_\alpha)$ and the action of the spin generator
$S_{\alpha\beta}$ is defined by [Gazeau, $1985$] \b
S_{\alpha\beta}K_\gamma=-i(\eta_{\alpha\gamma}K_{\beta}-\eta_{\beta\gamma}K_\alpha).\e
The operator $Q_1^{(1)}$ commutes with the action of the group
generators and consequently it is constant on each unitary
irreducible representation. In fact, the vector
UIR's can be classified by using the eigenvalues of $Q_1^{(1)}$, {\it
i.e.}, $<Q_1^{(1)}>$,
    \b (Q_1^{(1)}-<Q_1^{(1)}>)K(x)=0. \e
>From Takahashi [Takahashi, $1963$] we get the following classification
scheme:
    $$ Q^{(1)}=(-p(p+1)-(q+1)(q-2))I_d , $$
    $$ Q^{(2)}=(-p(p+1)q(q-1))I_d. $$
In the present context three types of vector UIR are distinguished for
$SO_0(1,4)$ according to the range of values of parameters $q$ and $p$.
[Dixmier, $1961$; Takahashi, $1963$], namely 
\newline 
i) the UIR's $U^{1,\nu}$ of the principal series, for which
 $$p=s=1,\;\; q=n+2=\frac{1}{2}+i\nu,$$          
 \begin{equation}
   <Q_1^{(1)}>=-n(n+3)-1(1+1)=\nu^2+\frac{1}{4},\;\;\;\nu \geq 0,
        \end{equation}
with parameter $\nu \in \R$. Note that
$U^{1,\nu}$ and $U^{1,-\nu}$ are equivalent.\newline
ii) the UIR's $V^{1,q}$ of the complementary series, for which
$$ p=s=1,$$
         \begin{equation}
   <Q_1^{(1)}>=q-q^2\equiv \mu,\;\;\;0<\mu <\frac{1}{4},
        \end{equation}
iii) the UIR's $\Pi^{\pm}_{p,1}$ of the discrete series, for which
            \b
   <Q_1^{(1)}>=-p(p+1)+2,\;\;\;p \geq 1,\;\;q=1.
        \e
In the ``massless'' case we have $p=q=s=1, \; \Pi^{\pm}_{1,1}$.

Using $(2.14)$ and $(2.15)$, the action $Q_1$ on the
five-component vector field $K$ gives
\b Q_1K(x)=
(Q_0-2)K(x)+2x\bar \partial\cdot K(x)-2\partial x\cdot K(x).\e
If the vector field satisfies the divergenceless condition
       \b \partial\cdot K(x)=\bar \partial\cdot K(x)=0, \e
it can be biunivocally associated with a UIR of the dS group. Therefore,
with the
conditions  $x\cdot K=0$ and $\bar \partial\cdot K=0$ and by using
Eq.($2.16$) and Eq.($2.20$), we
obtain
       \b (\Box_H+2H^2+H^2<Q_1>)K_\alpha(x)=0, \e
which has the same form as $(2.13)$. Comparing with the latter, we get
$H^2<Q_1>=m^2_H$. 
It follows the respective mass relations for the
three types of UIR:
   $$m^2_p=H^2(\nu^2+\frac{1}{4}),\;\;\;\nu\geq 0\;\;\mbox{(for the
principal
series)},$$
 $$m^2_c=H^2\mu,\;\;\;0<\mu< \frac{1}{4}\;\;\mbox{(for the complementary
series)},$$
        \b m^2_d=H^2(2-p(p+1)),\;\;\;p\geq 1\;\;\mbox{(for the discrete
series).}\e
 In particular, for the discrete series representation with $p=1$, the
``mass'' parameter is zero, and for $p>1$, it is purely imaginary. We
shall return to this point later. In this paper, we only consider the
``massive'' vector field, {\it i.e.} that one for which the values
assumed by the parameter $m_H$ correspond to the principal series
representations. Eq. $(2.22)$ then reads
      \b (\Box_H+2H^2+m^2_p)K_\alpha(x)=0.\e
Let us recall at this point the physical content of the principal series
representation from the
point of view of a Minkowskian observer at the limit $H=0$.
The principal series UIR $U^{1,\nu},\;\;\nu \geq 0$, contracts
toward the  direct sum of two vector massive Poincar\'e UIR's
$P^<(m,1)$ and $P^>(m,1)$, with negative and positive energies
respectively [Mickelsson, Nielderle, $1972$].
          \begin{equation}
    U^{1,\nu}  {H\rightarrow 0 \over \nu \rightarrow \infty }
\longrightarrow P^<(m,1)\bigoplus
P^>(m,1).
        \end{equation}
The contraction limit has to be understood through the constraint
$m= H\nu$.  The quantity $m_H$, supposed to
 depend on $H$, goes to the classical mass $m$ when the curvature goes
to zero, \newline
In contrast, only one representation in the discrete series with $p=1$
has a
Minkowskian interpretation. It was denoted by $(\Pi^{\pm}_{1,1})$ by
Dixmier
[Dixmier, $1961$]. The signs $\pm$ correspond to two types of helicity
for
the massless vector field. The representation  $\Pi^+_{1,1}$ has a
unique extension to a direct
 sum of two UIR's $C(2;1,0)$ and $C(-2;1,0)$ of the conformal group
$SO_0(2,4)$ 
with positive and negative energies respectively [Barut, B\"ohm, $1970$,
Angelopoulos, Laoues
$1998$]. The latter restrict to the vector
   massless Poincar\'e RUI's $P^>(0, 1)$ and $P^<(0,1)$ with positive
and negative energies
respectively.
    The following diagrams illustrate these connections
\b \left. \begin{array}{ccccccc}
     &             & {\cal C}(2,1,0)
& &{\cal C}(2,1,0)   &\hookleftarrow &{\cal P}^{>}(0,1)\\
 \Pi^+_{1,1} &\hookrightarrow  & \oplus            
&\stackrel{H=0}{\longrightarrow} & \oplus  & &\oplus  \\
     &             & {\cal C}(-2,1,0)&
& {\cal C}(-2,1,0)  &\hookleftarrow &{\cal P}^{<}(0,1),\\
    \end{array} \right. \e
 \b \left. \begin{array}{ccccccc}
     &             & {\cal C}(2,0,1)
& &{\cal C}(2,0,1) &\hookleftarrow &{\cal P}^{>}(0,-1)\\
 \Pi^-_{1,1} &\hookrightarrow  & \oplus
&\stackrel{H=0}{\longrightarrow} &  \oplus & &\oplus  \\
     &             & {\cal C}(-2,0,1)&
& {\cal C}(-2,0,1)   &\hookleftarrow &{\cal       P}^{<}(0,-1),\\
    \end{array} \right. \e
where the arrows $\hookrightarrow $ designate unique extension
and $ {\cal P}^{ \stackrel{>}{<}}(0,1)$ are the massless
Poincar\'e UIR with  positive and negative energies and 
positive helicity. $ {\cal P}^{ \stackrel{>}{<}}(0,-1)$ are the
massless Poincar\'e UIR with positive and negative energies and
negative helicity.
Finally, all
other representations have no non-ambiguous Minkowskian counterpart.

\setcounter{equation}{0}
\section{dS-vector plane waves}

In the five-component vector field notation $K_\alpha(x)$, the solution
can be written 
in terms of the scalar fields. More precisely we put [Gazeau, Hans,
$1988$]
     \b K_\alpha (x)=\bar Z_\alpha \phi_1+D_{1\alpha} \phi_2, \e
where $Z$ is a constant vector $(\bar Z_\alpha
=\theta_{\alpha\beta} Z^\beta=Z_\alpha+H^2x_\alpha x\cdot Z,\;x\cdot
\bar
Z=0$) and $D_{1\alpha}= H^{-2}\bar \partial_\alpha$ is the
generalized gradient. An arbitrary five-component vector $Z_\alpha$
 is obtained in the same way for an arbitrary four-component spinor 
[Gazeau and al, $1999$-b]. We choose $Z_\alpha$
such that at the limit $H=0$, one obtains the vector field in the
Minkowskian space. In this limit, $Z_\alpha$ must be related in some
sense with the
usual massive polarization vectors. There are three polarization vectors
for the
dS vector field $(s=1)$. They generate the  vector representation
of the group $SU(2)\;\;(2s+1=3)$.

Putting $K_\alpha$
in $(2.16)$ and using the following relations
   \b  Q_1D_1\phi_2=D_1Q_0\phi_2 ,\e
     \b Q_1\bar Z_{\alpha}\phi_1=\bar
           Z_{\alpha}(Q_0-2)\phi_1-2H^2D_1(x\cdot Z)\phi_1,\e
we find that the scalar fields $\phi_1$ and $\phi_2$ must obey :
    \b  ( Q_0-(\nu^2+\frac{9}{4}))\phi_1=0=( 
\Box_H+H^2(\nu^2+\frac{9}{4}))\phi_1,\e
 \b Q_0\phi_2-(\nu^2+\frac{1}{4})\phi_2-2H^2(x.Z)\phi_1=0 .\e
So $\phi_1$ is a ``massive'' scalar field (principal series). Now
the vector field must satisfy the divergencelessness condition $(2.21)$.
Therefore
we have from $(3.1)$ $$ \partial\cdot K(x)=0 \; \Rightarrow
\;Q_0\phi_2=Z\cdot \bar \partial
\phi_1+4H^2Z\cdot x\phi_1.$$ 
We have here used the relation
   $$ \partial \cdot \bar Z \phi=Z\cdot \bar \partial \phi+4H^2Z\cdot
x\phi.$$
So the field $\phi_2$ can be written in terms of $\phi_1$ 
    \b \phi_2 =\frac{1}{\nu^2+\frac{1}{4}} [Z\cdot \bar \partial
\phi_1+2H^2x\cdot Z\phi_1].\e
Eq.$(3.4)$ has solutions which are  homogeneous with degree
$\sigma=-\frac{3}{2} \pm i\nu$, and which are identified as dS plane
waves [Bros, Gazeau, Moschella $1994$]
  \b \phi_1(x)=(Hx\cdot \xi)^\sigma ,\e
where $\xi \in \R^5 $ lies on the null cone
  $ {\cal C} = \{ \xi \in \R^5;\;\; \xi^2=0 \}$. It follows that the two
possible solutions   for $K$ are
  $$ K_{1\alpha}(x)=[\bar Z_{\alpha}+\frac{1}{\nu^2+\frac{1}{4}}
D_{1\alpha}(Z\cdot \bar \partial+2H^2x\cdot Z)](Hx\cdot
\xi)^{-\frac{3}{2}+i\nu}$$
    \b  \equiv{\cal E}_{1\alpha}(x,\xi,Z)(Hx\cdot
\xi)^{-\frac{3}{2}+i\nu},\e
$$ K_{2\alpha}(x)=[\bar Z_{\alpha}+\frac{1}{\nu^2+\frac{1}{4}}
D_{1\alpha}(Z\cdot \bar \partial
+2H^2x\cdot Z)](Hx\cdot \xi)^{-\frac{3}{2}-i\nu}$$         
 \begin{equation}
\equiv{\cal
E}_{2\alpha}(x,\xi,Z)(Hx\cdot\xi)^{-\frac{3}{2}-i\nu},\end{equation}
where ${\cal E}_{1\alpha}$ are the generalized polarization vector
and ${\cal E}_{2\alpha}^*={\cal E}_{1\alpha} \equiv {\cal
E}_{\alpha}$. The generalized polarization vector ${\cal
E}_{\alpha}(x,\xi,Z)$  is  function of the  space-time point $x$.
Its expression is given by $$ {\cal
E}_{\alpha}(x,\xi,Z)=(\frac{3}{2}-i\nu) \bar
Z_{\alpha}$$
    \begin{equation}  +\frac{1}{\nu^2+\frac{1}{4}}
    \left[(i\nu-\frac{3}{2})(i\nu-\frac{5}{2}) \frac{\bar Z\cdot
\xi}{(Hx\cdot \xi)^2}
     +3(i\nu-\frac{3}{2}) \frac{Z\cdot x}{x\cdot \xi} \right]\bar
\xi_{\alpha} ,\end{equation}
in which  $\bar \xi_{\alpha}=\theta_{\alpha\beta}\xi^\beta$.  In the
limit $H =0$, $(Hx\cdot \xi)^{-\frac{3}{2}-i\nu}$ and ${\cal
E}_{\alpha}(x,\xi,Z)$ 
behave like the plane wave $e^{ ik\cdot X}$ and the
polarization vector in the Minkowski space respectively. If we
parametrize $\xi$
in terms of the four-momentum of the limit Minkowskian particle
of mass $m$,
 \b \xi=(\frac{k^0}{mc}=\sqrt{\frac{\vec k^2}{m^2c^2}+1},\frac{\vec
k}{mc},-1),\e
we have from $(3.8)$
   $$ \lim_{H \rightarrow 0}(Hx(X)\cdot \xi)^{-\frac{3}{2}+i\nu}{\cal
E}_{\alpha}(x,\xi,Z) =
     \left(Z_{\mu}-\frac{Z_{\nu}k^{\nu}}{m^2}k_{\mu}\right)e^{ ik\cdot
X}$$
    \begin{equation}  \equiv \epsilon^{(\lambda)}_{\mu}(k)e^{ ik\cdot
X},
      \;\lambda=1,2,3.\end{equation}
Here, the dS point $x=x_H(X)$ has been expressed in terms the
Minkowskian variable $X=(X_0=ct, \vec X)$ measured in units of the
dS radius $H^{-1}$:
 $$ x_H(X)=(x^0=H^{-1}\sinh HX^0, \vec x=H^{-1} \frac{\vec X}{\parallel
\vec X\parallel}
  \cosh HX^0 \sin H\parallel \vec X\parallel,$$
\b x^4= H^{-1}  \cosh HX^0 \cos H\parallel \vec X\parallel).\e
Note that $(X^0, \vec X)$ are global coordinates. The
compact spherical nature of space at fixed $X^0$ is apparent
in $(3.13)$. The $\epsilon^{(\lambda)}_{\mu}(k)$'s are the three
polarization vectors in the Minkowski space [Itzykson, Zuber,
$1982$]:       \begin{equation}  \epsilon^{(\lambda)}\cdot k=0,\;\;
\epsilon^{(\lambda)}\cdot \epsilon^{(\lambda')}=\delta_{\lambda
\lambda'}, \end{equation}
\begin{equation} \sum_{\lambda =1}^3\epsilon^{(\lambda)}_{\mu}(k)
\epsilon^{(\lambda)}_{\nu}(k)=    -(\eta_{\mu
\nu}-\frac{k_{\mu}k_{\nu}}{m^2}) ,\end{equation}
and $\eta_{\mu \nu}=\mbox{diag}(1,-1,-1,-1)$.
For simplicity we choose three five-component vectors $Z^{\lambda}$
which obey the transverse
constraints:
$$ Z^{(\lambda)}\cdot \xi=0,$$
     \b Z^{(\lambda)}_{\alpha}=(\epsilon^{(\lambda)}_{\mu}(k),
         Z^{(\lambda)}_4=0).\e
The generalized polarization vectors now read:
    $$ {\cal E}_{\alpha}^{(\lambda)}(x,\xi) \equiv {\cal E
}_{\alpha}(x,\xi,Z^{(\lambda)})$$
   \begin{equation}    =
(\frac{3}{2} -i\nu)\bar
Z_{\alpha}^{(\lambda)}            
+(i\nu-\frac{3}{2})(-i\nu+\frac{1}{2})          
\frac{Z^{(\lambda)}\cdot x}{x\cdot \xi} \bar \xi_{\alpha}.\end{equation} 
Finally, the two solutions
for the dS-vector field take the form:
\b K_{1\alpha}(x)={\cal E}_{\alpha}^{(\lambda)}(x,\xi)(Hx\cdot
\xi)^{-\frac{3}{2}+i\nu},\e
   \begin{equation} K_{2\alpha}(x)=
{\cal E}_{\alpha}^{*(\lambda)}(x,\xi)(Hx\cdot
\xi)^{-\frac{3}{2}-i\nu},\end{equation}
where ${\cal E}_{\alpha}^{*(\lambda)}$ is given by $(3.17)$.
These solutions are not globally defined due to the ambiguity on the
phase factor. For a complete determination, one may consider the
solution
in the complex de Sitter space $X_H^{(c)}$ [Bros, Moschella;
$1996$]
       \b K_{1\alpha}(z)={\cal E}_{\alpha}^{(\lambda)}(z,\xi)(Hz\cdot
\xi)^{-\frac{3}{2}+i\nu},\e
   \begin{equation} K_{2\alpha}(z)={\cal E}_{\alpha}^{*(\lambda)}(z,\xi)
(Hz\cdot \xi)^{-\frac{3}{2}-i\nu},\end{equation}
in which $z \in X_H^{(c)}=\{ z=x+iy\in  \C^5;\;\;\eta_{\alpha
\beta}z^\alpha z^\beta= (z^0)^2-\vec z.\vec
z-(z^4)^2=-H^{-2}\}$.

In the same way as in the Minkowskian space, it is seen that for the
scalar 
and vector fields the two solutions (3.18) and (3.19) are  complex
conjugate of each other. On the other hand, for the spinor field there
is
no such relation between them [Gazeau and al $1999$-b].

\setcounter{equation}{0}
\section{Two-point function and quantum field}

We here follow the procedure already presented and discussed in previous
works. Let us briefly
recall
 the required conditions on the matrix Wightman two-point function
${\cal
W}(x,x')$.  Its matrix elements ${\cal W}_{\alpha \alpha'}$ are
defined by
     \begin{equation}  {\cal W}_{\alpha \alpha'}(x,x')=
     \langle \Omega,K_{\alpha}(x)K_{\alpha'}(x')\Omega \rangle,\ 
\alpha, \,  \alpha'=0,\cdots
,4    
\end{equation} where $x,x'\in X_H$. These functions entirely  encode
 the theory of the generalized free fields on dS
space-time $X_H$. They have to satisfy the following
requirements:
\begin{enumerate}
\item[a)] {\bf Positiveness}

for any test function
$f_\alpha \in {\cal D}(X_H)$, we have
\begin{equation} \int _{X_H \times X_H}  f^{*\alpha}(x){\cal W}_{\alpha
\alpha'}(x,x')
f^{\alpha'}(x')d\sigma(x)d\sigma(x')\geq       0,\end{equation}
where $ f^*$ is the complex conjugate of $f$ and $d\sigma (x)$
denotes the dS-invariant measure on $X_H$ [Bros, Moschella,
$1996$]. ${\cal D}(X_H)$ is the space of function $C^\infty$ with
compact support in $X_H$,
\item[b)] {\bf Locality} 

for every
space-like separated pair $(x,x')$, {\it i.e.} $x\cdot x'>-H^{-2}$,
\begin{equation}{\cal W}_{\alpha \alpha'}(x,x')={\cal W}_{ \alpha'
\alpha}(x',x),\end{equation}
 \item[c)] {\bf Covariance}
     \begin{equation}g^{-1}{\cal W} (g x,g x')g={\cal
W}(x,x'),\end{equation}
     where $g\in SO_0(1,4)$,
\item[d)] {\bf Normal analyticity}

 ${\cal W}_{\alpha \alpha'
}(x,x')$ is the boundary value (in the distributional sense ) of
an analytic function $W_{\alpha \alpha'}(z,z')$,
\item[e)] {\bf Transversality}
\begin{equation} x\cdot {\cal W}(x,x')=0=x'\cdot {\cal
W}(x,x'),\end{equation}
\item[f)] {\bf Divergencelessness}
\begin{equation} \partial_x\cdot {\cal W}(x,x')=0=\partial_{x'}\cdot
{\cal W}(x,x').
\end{equation}
\end{enumerate}
 As it has been comprehensively justified by the theorem $4.1$ of Ref.
[Bros, Moschella, $1996$],
the analytic two-point function
$W_{\alpha \alpha'}^{\nu}(z,z')$ is obtained from the complexified plane
waves of the type $(3.20)$ and $(3.21)$
\begin{equation} K_1^{\xi,\lambda}(z)=(Hz\cdot \xi)^{-\frac{3}{2}+ i
\nu}
 {\cal E}^{\lambda}(z,\xi),\end{equation}
   \begin{equation}K_2^{\xi,\lambda}(z)=(Hz\cdot \xi)^{-\frac{3}{2}- i
\nu}{\cal
E}^{*\lambda}(z,\xi),\end{equation} where $ {\cal E}^{\lambda}(z,\xi)$
is defined by $(3.17)$. 
Explicitely, it is given in terms of the following class of integral
representations
   \begin{equation} W^\nu _{\alpha \alpha'}(z,z')= c_\nu \int_T (z\cdot
\xi)^{-\frac{3}{2}-i\nu}
   (\xi\cdot z')^{-\frac{3}{2}+i\nu}\sum_{\lambda = 1}^3 {\cal    
E}^{\lambda}_{\alpha }(z,\xi)
    {\cal E}^{*\lambda}_{\alpha'}              
(z',\xi)d\mu_T(\xi).\end{equation}
Here $T$ denotes the orbital basis of ${\cal C}^+=\{\xi \in {\cal
C};\; \xi^0>0\}$. $d\mu_T(\xi)$ is an invariant measure 
defined by
 \b d\mu_T(\xi)=i_\Xi w_{{\cal
C}^+}\mid_T,\e where $i_\Xi w_{{\cal C}^+}$ denotes the $3$-form on
${\cal C}^+$ obtained from the
contraction of the vector field
$\Xi$ with the volume form [Bros, Moschella, $1996$]
 \b w_{{\cal
C}^+}=\frac{d\xi^0\wedge \cdots \wedge d\xi^4 }{d(\xi\cdot \xi)}.\e
The coefficient $c_\nu$ is a normalization constant which is fixed by
local Hadamard condition. The latter selects a unique vacuum state for
quantum vector fields which satisfy the dS field equation.

The functions ${\cal
W}^{\nu}_{\alpha \alpha'}(x,x')$, which are
solution to the wave equation $(2.16)$, can be found simply in 
terms of scalar Wightman two-point
functions ${\cal W}_i^{\nu}(x,x'), i= 1,2$,
 without resorting to any explicit calculation of the
integral $(4.9)$. By using the recurrence formula
$(3.1)$ we obtain
\begin{equation}{\cal W}^{\nu}_{\alpha \alpha'}(x,x')=\theta_{\alpha}.
\theta'_{\alpha' }{\cal W}_1^{\nu}(x,x')+H^{-2}\bar
\partial_{\alpha}\bar \partial'_{\alpha'} {\cal
W}_2^{\nu}(x,x').\end{equation}
By prescribing ${\cal W}^{\nu}_{\alpha \alpha'}$ to obey Eq. $(2.16)$
and by using the previous
conditions and relations $(3.2)$ and $(3.3)$, it is found that ${\cal
W}_1$ satisfies the
equation:
  \begin{equation}   [Q_0-(\nu^2+\frac{9}{4})]{\cal  
W}_1^{\nu}(x,x')=0,\end{equation}
whilst ${\cal W}_2$ is given in terms of ${\cal W}_1$ by
  \begin{equation} {\cal W}_2^{\nu}(x,x')=\frac{1}{\nu^2+\frac{1}{4}}
  [H^{-2}\bar \partial\cdot \bar \partial'{\cal W}_1^{\nu}(x,x')+2H^2
x\cdot x'{\cal  
W}_1^{\nu}(x,x')].\end{equation} 
The vector Wightman function can then be written in the form:
  \begin{equation} {\cal W}^{\nu}_{\alpha \alpha'}(x,x')=
D_{\alpha \alpha'}(x,\bar \partial;x',\bar \partial')
  {\cal W}_1^{\nu}(x,x'),\end{equation}
where ${\cal W}_1^{\nu}$ is solution to $(4.13)$ and
  \b D_{\alpha \alpha'}=\theta_{\alpha}\cdot \theta'_{\alpha' }
  +\frac{1}{H^2(\nu^2+\frac{1}{4})}\bar \partial_{\alpha}\bar
    \partial'_{\alpha'}[H^{-2}\bar \partial\cdot \bar \partial'+2H^2
x\cdot x'].\e
At the limit $H=0$, the corresponding vector Wightman two-point function
in
Minkowski space is obtained in terms of the
Wightman two-point function ${\cal W}^P(X,X')$ for the scalar field in 
the Minkowski
space.
 \b {\cal W}_{\mu \nu}(X,X')=\left[ \eta_{\mu
\nu}+\frac{1}{m^2} \frac{\partial^2} {\partial X^\mu \partial
X^\nu} \right]{\cal W}^P(X,X').\e 
For the analytic function $W_1^{\nu}$  we recall that we have the
following expression
[Bros, Moschella,
$1996$]
   \b W_1^{\nu}(z,z')= C_\nu P_{-\frac{3}{2}+i\nu}^{(5)}(H^2z\cdot
z'),\e
where $C_{\nu}=2\pi^2 e^{\pi \nu}H^3 c_\nu$ and \b
c_\nu=\frac{e^{-\pi
\nu}\Gamma(\frac{3}{2}+i\nu)\Gamma(\frac{3}{2}-i\nu)}{2^5\pi^4H}.\e
$P_{-\frac{3}{2}+i\nu}^{(5)}$ is the generalized Legendre function
of the first kind. Finally we get the analytic function  $W_{\alpha
\alpha'}(z,z')$ in term of the latter: 
 \b  W^{\nu}_{\alpha
\alpha'}(z,z')=C_{\nu}D_{\alpha \alpha'}(z,\bar \partial; z',\bar
\partial')P_{-\frac{3}{2}+i\nu}^{(5)}(H^2z\cdot z').\e 
\noindent Its analyticity properties follow 
 from the expression of the plane-waves $(3.20)$ and $(3.21)$.

The positiveness property is issued from the hermiticity
condition. The proof makes use of the Fourier-Bros transformation on
$X_H$
[Bros, Moschella, $1996$]. The hermiticity property is also
obtained by considering boundary values of the following identity

  \b W_{\alpha \alpha'}(z,z')= W_{\alpha' \alpha}^*( z'^*, z^*),\e
\noindent which is easily checked on Eq. $(4.9)$.

The relation $ g^{-1}z_1\cdot \xi=z_1\cdot g\xi$  and the
independence of the integral $(4.9)$ with respect to 
the selected orbital basis $T$ entail the
covariance property
\begin{equation} g^{-1}{\cal W} (g x,gx')g={\cal W}(x,x').\end{equation}
In order to prove the locality condition, the following relation 
is needed [Bros, Moschella $1996$]
    \b P_{-\frac{3}{2}+i\nu}^{(5)}(H^2z_1\cdot z_2) =
P_{-\frac{3}{2}-i\nu}^{(5)}(H^2z_1\cdot z_2).\e 
It follows the hermiticity
  \b W^\nu_{\alpha \alpha'} (z,z')= W^\nu_{\alpha' \alpha }
(z^{'*},z^*).\e
It is noted that the space-like separated pair ($x,x'$)  lies in the
same orbit of the complex  dS group as the pairs ($z,z'$) and
($z'^{*},z^*$)  [Bros, Moschella, $1996$], and so the
locality condition ${\cal W}_{\alpha \alpha'}(x,x')={\cal
W}_{\alpha' \alpha}(x',x)$ holds for the former.

Now, going back to Eq. $(4.9)$, the boundary value of $W^\nu (z,z')$
gives rise to 
the following integral
representation of the Wightman two-point function itself:
 $$ {\cal W}_{\alpha \alpha'}^\nu (x,x')=c_{\nu}\int_T [(x\cdot
\xi)_+^{-\frac{3}{2}-i\nu}+
 e^{i\pi(-\frac{3}{2}-i\nu)}       (x\cdot \xi)_-^{-\frac{3}{2}-i\nu}]$$
\b[(x'\cdot
\xi)_+^{-\frac{3}{2}+i\nu}+e^{-i\pi(-\frac{3}{2}+i\nu)}(x'\cdot
\xi)_-^{-\frac{3}{2}
+i\nu}] \sum_{\lambda = 1}^3 {\cal  E}^{\lambda}_{\alpha }(x,\xi)
 {\cal E}^{*\lambda}_{\alpha'} (x',\xi) d\mu_T.\e
where $ (x\cdot \xi)_+=\left\{\begin{array}{clcr} 0 & \mbox{for} \;
x\cdot  
\xi\leq 0\\ (x\cdot \xi) & \mbox{for} \;x\cdot \xi>0. \\ \end{array}
\right.$ [Gel'fand, Shilov, $1964$]. This relation 
defines the two-point  function in term of global plane waves on
$X_H$. 

The explicit knowledge of ${\cal W}$ allows
us to make the QF formalism work. The vector field $K(x)$ is
expected to be an operator-valued distribution on $X_H$ acting on a
Hilbert space ${\cal H}$. In terms of Hilbert space and
field-operators the properties of the Wightman two-point
functions are equivalent to the following  conditions [Streater,
Wightman, $1964$]:
\begin{enumerate}

\item {\bf Existence of an unitary irreducible representation of
the dS group}
 $$ U^{1,\nu};\;\; V^{1,q};\;\; \Pi^\pm_{p,1}, \;
p\neq 1,$$
\item {\bf Existence of a Hilbert space} ${\cal H}$ 

with positive
definite metric that can be described as the Hilbertian sum \b
{\cal H}={\cal H}_0\bigoplus[\bigoplus_{n=1}^{\infty}S{\cal
H}_1^{\bigotimes n}],\e 
where $S$ denotes the symmetrization operation
and ${\cal H}_0=\{ \lambda \Omega,\;\; \lambda \in \C\}$. ${\cal
H}_1$ is precisely equipped with the scalar product \b
(h_1,h_2)=\int_{X_H
\times X_H} h_1^{*\alpha}(x){\cal W}_{\alpha
\alpha'}(x,x')h_2^{\alpha'}(x')d\sigma(x)d\sigma(x')\geq 0,\e
where $h_\alpha \in {\cal D}(X_H)$,
\item {\bf Existence of at least one ``vacuum state''}  $\Omega$,

cyclic for the polynomial algebra of field operators and invariant
under the representation of dS group.
\item {\bf Covariance } 

of the field operators under the
representation of dS group,
\item{\bf Locality} 

for every space-like separated pair $(x,x')$
\b [K _\alpha(x) , K_{\alpha}(x') ]=0.\e
\item {\bf KMS condition or geodesic spectral condition} [Bros,
Moschella, $1996$] 

which means that the vacuum is defined as a physical
state with the temperature $T=\frac{H}{2\pi}$,
\item {\bf Transversality}
\begin{equation} x\cdot K(x)=0,\end{equation}
\item {\bf Divergencelessness}
\begin{equation} \partial\cdot K(x)=0.  \end{equation}
\end{enumerate}

In terms of annihilation and creation operators, the field operator
 $ K(f)=K^+ (f) +K^- (f)$ is defined by
$$(K^-(f)h)^{(n)}(\alpha_1, x_1;\alpha_2, x_2;\cdots
;\alpha_n,x_n)=\sqrt{n+1 }$$
   \begin{equation}\int_{ X_H \times X_H } f_{\alpha}(x) {\cal W}^{\beta
\alpha }
   (y,x)h^{(n+1)}(\beta, y;\alpha_1, x_1;\cdots ;\alpha_n,x_n)d\sigma
(x)d\sigma (y)\end{equation}
 $$(K^+(f)h)^{(n)}(\alpha_1, x_1;\alpha_2, x_2;\cdots ;\alpha_n,x_n)=$$
\begin{equation}\frac{1}{\sqrt{n}}\sum_{k=1}^n  f_{\alpha_k}(x_k)
 h^{(n-1)}(\alpha_1, x_1;....;\hat{\alpha_k}, \hat{x_k};....;\alpha_n,
x_n),
     \end{equation}
in which symbols with hat are omitted . Using the Fourier-Bros
transformation on $X_H$, the unsmeared operators
$K(x)$ can be written as
 $$ K(x)=\int_T\sum_{\lambda=1}^3 \{\;
a_{\lambda}(\xi,\nu){\cal
E}^{\lambda}(x,\xi)[(x\cdot \xi)_+^{-\frac{3}{2}-i\nu}+
e^{i\pi(-\frac{3}{2}-i\nu)}(x\cdot \xi)_-^{-\frac{3}{2}-i\nu}] $$ \b
+a^{\dag}_{\lambda}(\xi,\nu){\cal
E}^{*\lambda}(x,\xi)[(x\cdot \xi)_+^{-\frac{3}{2}+i\nu}+
e^{-i\pi(-\frac{3}{2}+i\nu)}(x\cdot \xi)_-^{-\frac{3}{2}+i\nu}]\; \}
d\mu_T(\xi),\e where $a_{\lambda}(\xi,\nu)$ is defined by
       \b a_{\lambda}(\xi,\nu)|\Omega>=0,\e
The integral representation $(4.33)$ is independent of 
the orbital basis $T$ if the following relation exists
 $$ \;a_{\lambda}(l
\xi,\nu)=l ^{-\frac{3}{2}+i\nu}a_{\lambda}(\xi,\nu).$$ 
The number operator $N$ is defined as \b N^{(\lambda)}=\int_T
d\mu_T(\xi)
a_{\lambda}^{\dag}(\xi,\nu)a_{\lambda}(\xi,\nu).\e 
This integral
is also independent of the orbital basis $T$. A
``one-particle'' state is defined via the ``creation'' operator in a
Fock
space \b a_{\lambda}^{\dag}(\xi,\nu)|\Omega>=|\xi,\lambda>.\e 
So far the
physical meaning of $N$ and the states $|\xi,\lambda>$ have not been
clarified. Let us work with the hyperbolic-type submanifold $T_4=T_4^+
\cup T_4^-$ defined by
 \b T^\pm_4=\{ \xi \in {\cal C}^+;\;\; \xi^4=\pm1\}.\e
In this orbital basis we have \b
[a_{\lambda}(\xi,\nu),a_{\lambda'}^{\dag}(\xi',\nu)]=
c_{\nu}\delta_{\lambda\lambda'}\frac{\xi^0}{|\xi^4|}\delta^3(\vec
\xi-\vec \xi'),\e or \b
<\Omega,a_{\lambda}(\xi,\nu)a_{\lambda'}^{\dag}(\xi',\nu)\Omega>=
c_{\nu}\delta_{\lambda\lambda'}\frac{\xi^0}{|\xi^4|}\delta^3(\vec
\xi-\vec  \xi').\e 
The relation between the
quantum field in dS and its Minkowskian counterpart has now become
apparent. 
In the limit $H=0$,
the equation
$(4.33)$ goes to the corresponding  massive vector field expansion in 
Minkowski space-time.

\section{Conclusion}

In this paper, we have considered the ``massive'' vector field 
associated to the principal series of the dS group $SO_0(1,4)$ with
$<Q_\nu>=\nu^2+\frac{1}{4},\;\;\nu \geq 0$ and with corresponding
``mass''
$m_p^2=H^2(\nu^2+\frac{1}{4})$. For the complementary series
($<Q_\mu>=\mu,\;\;0<\mu <\frac{1}{4}$) and the discrete  series
($<Q_p>=2-p(p+1),\;\;p \geq 1$), we can replace $\nu$ respectively
by $\pm \sqrt{\mu-\frac{1}{4}}$ and $\pm
\sqrt{\frac{7}{4}-p(p+1)}$.

In the case of the
complementary series the associated ``mass'' is positive
($m_c^2=H^2\mu,\;\;0<\mu <\frac{1}{4}$), but in the limit $H=0$
there are no physically meaningful representation of the Poincar\`e
group. So the physical meaning of these fields is not clear yet.

For the discrete series the associated ``mass'' is zero or  imaginary
($m_d^2=H^2\{2-p(p+1)\},\;\;p \geq 1$). Only one among the discrete
series
representations, namely that one corresponding to 
$p=1$ has a physically meaningful Poincar\'e limit. The latter is
precisely 
the ``massless'' vector field (QED in dS space) and $\nu$ must be
replaced by $\pm
\frac{i}{2}$ in the previous formulas. Yet the generalized
polarization vector ${\cal E}$ (Eq. $(3.10)$) and the scalar field
$\phi_2$ (Eq. $(3.6)$)  diverge at the limit. This type of singularity
is actually due to the divergencelessness condition for
associating this field with a specific UIR of the dS group. It can be 
as
well understood from the equation allowing to determine $\phi_2$
in terms of $\phi_1$. To solve this problem, the divergencelessness
condition must be dropped out.
Then the vector field is associated with an indecomposable
representation of the dS group. This
situation will be considered in a forthcoming paper.

\end{document}